\documentclass[aps,amsmath,prb,superscriptaddress,showpacs,floatfix,twocolumn]{revtex4}
\usepackage{graphicx}
\usepackage{amsmath}
\usepackage{amsfonts}
\usepackage{amssymb}
\usepackage{dcolumn}
\usepackage{natbib}
\usepackage[papersize={8.5in,11in}]{geometry}

\begin{document}

\title{Thermal hopping and retrapping of a Brownian particle in the tilted periodic potential of a NbN/MgO/NbN Josephson junction}

\author{Luigi Longobardi}
\email{llongobardi@ms.cc.sunysb.edu}
\affiliation{Seconda Universit\'{a} degli Studi di Napoli, Dipartimento di Ingegneria dell'Informazione, via Roma 29, 81031 Aversa (Ce) Italy}
\affiliation{CNR-SPIN UOS Napoli, Complesso Universitario di Monte Sant'Angelo via Cinthia, 80126 Napoli (Na) Italy}
\author{Davide Massarotti}
\affiliation{Universit\'{a} degli Studi di Napoli "Federico II", Dipartimento di Scienze Fisiche, via Cinthia, 80126 Napoli (Na) Italy}
\affiliation{CNR-SPIN UOS Napoli, Complesso Universitario di Monte Sant'Angelo via Cinthia, 80126 Napoli (Na) Italy}
\author{Giacomo Rotoli}
\affiliation{Seconda Universit\'{a} degli Studi di Napoli, Dipartimento di Ingegneria dell'Informazione, via Roma 29, 81031 Aversa (Ce) Italy}
\author{Daniela Stornaiuolo}
\affiliation{CNR-SPIN UOS Napoli, Complesso Universitario di Monte Sant'Angelo via Cinthia, 80126 Napoli (Na) Italy}
\author{Gianpaolo Papari}
\affiliation{NEST, CNR-NANO, and Scuola Normale Superiore di Pisa , Piazza dei Cavalieri  7, 56126 Pisa, Italy}
\author{Akira Kawakami}
\affiliation{Advanced ICT Research Institute, National Institute of Information and Communications Technology, 588-2 Iwaoka-cho, Nishi-ku, Kobe 651-2492, Japan}
\author{Giovanni Piero Pepe}
\affiliation{Universit\'{a} degli Studi di Napoli "Federico II", Dipartimento di Scienze Fisiche, via Cinthia, 80126 Napoli (Na) Italy}
\affiliation{CNR-SPIN UOS Napoli, Complesso Universitario di Monte Sant'Angelo via Cinthia, 80126 Napoli (Na) Italy}
\author{Antonio Barone}
\affiliation{Universit\'{a} degli Studi di Napoli "Federico II", Dipartimento di Scienze Fisiche, via Cinthia, 80126 Napoli (Na) Italy}
\affiliation{CNR-SPIN UOS Napoli, Complesso Universitario di Monte Sant'Angelo via Cinthia, 80126 Napoli (Na) Italy}
\author{Francesco Tafuri}
\affiliation{Seconda Universit\'{a} degli Studi di Napoli, Dipartimento di Ingegneria dell'Informazione, via Roma 29, 81031 Aversa (Ce) Italy}
\affiliation{CNR-SPIN UOS Napoli, Complesso Universitario di Monte Sant'Angelo via Cinthia, 80126 Napoli (Na) Italy}

\date{\today}

\begin{abstract}
We report on the occurrence of multiple hopping and retrapping of a Brownian particle in a tilted washboard potential. The escape dynamic has been studied experimentally by measuring the switching current distributions as a function of temperature in a moderately damped NbN/MgO/NbN Josephson junction.
At low temperatures the second moment of the distribution increases in agreement with calculations based on Kramers thermal activation regime. After a turn-over temperature $T^*$, the shape of the distributions starts changing and width decreases with temperature. We analyze the data through fit of the switching probability and Monte Carlo simulations and we find a good agreement with a model based on a multiple retrapping process.
\end{abstract}

\pacs{05.40.Jc, 74.50.+r, 85.25.Cp}

\maketitle

\section{Introduction}

Research on superconducting quantum systems with potentials for qubits applications has boosted  interest on several complementary aspects of coherence and dissipation. There is a growing evidence of the occurrence of a moderately damped regime (MDR) in superconducting Josephson junctions (JJs)  of various materials\cite{kautz90,kivioja2005,mannik2005,krasnov2005,fenton2008,Bae2009}. If we use the $Q=\omega_p R C$ parameter as a measure of dissipation in a junction \cite{barone,likharevbook}, where $\omega_p=(2eI_c/\hbar C)^{1/2}$ represents the plasma frequency at zero bias current, $I_c$ is the junction critical current and C and R are the junction capacitance and resistance,
a MDR  is present for  $1 < Q < 5$. This regime is quite distinct from the well-known case of underdamped systems (Q $>10$) \cite{devoret1985,martinis1987}, and apparently quite common in junctions characterized by low $I_c$. In view of a more and more relevant use of nanotechnologies in quantum superconducting electronics and therefore of low values of $I_c$, studies on MDR can offer novel insights on dissipative effects on Josephson junctions, and inspire appropriate designs to respond to specific circuit requirements.

Our analysis is based on measurements on low critical current density ($J_c$) NbN/MgO/NbN junctions. NbN is a material of great interest for sensor applications, as documented by several works both on junctions and nanowires\cite{delacour,Verevkin,ejrnaes09,marsili11}, and
it guarantees both fast non equilibrium electron-phonon relaxation times $\tau < 10 ps$ and higher gap values, when compared with traditional junction technologies based on Nb,
Al and Pb\cite{Ivlev_2005}. Low-$J_c$ NbN/MgO/NbN devices may contribute to set a more comprehensive NbN platform, and constitute a non-trivial extension to thicker barriers of the more established high-$J_c$ NbN junctions, usually designed for superconducting digital circuits \cite{Yamamori,Larrey}.

In this work we demonstrate the occurrence of a phase diffusion regime (PDR) induced by low-$J_c$ in large area junctions. Areas are one or two order of magnitudes larger than those of junctions where PDR has been previously observed\cite{kivioja2005,mannik2005}, and Josephson ($E_J$) and Coulomb ($E_C$) energies  entering in the tunneling process will scale with size in a different manner because of the diverse $J_C$ and specific capacitance ($C_s$) values. PDR prevails over thermal activation at temperatures above a threshold  $T^*$ of about 1.5K . Experimental results are very close to theoretical predictions\cite{kautz90,fenton2008} such to provide a reliable estimation of Q = 2.7.  We compare our results with the model of Fenton and Warburton (FW)\cite{fenton2008} which condenses ideas on phase diffusion of the last 20 years, offering a reliable methodology to evaluate levels of dissipation in MDR, in analogy with what well established for underdamped junctions\cite{devoret1985,martinis1987}. In particular FW model provides additional criteria to study low Q junctions to prove their MDR nature based on the asymmetry of the switching current histograms or its skewness $\gamma$, i.e. the ratio $m_3/\sigma^3$ where $m_3$ is the third central moment of the distribution, which are closely followed by experimental data. A physical picture emerges of moderately damped junctions, with a damping substantially independent of the frequency  and able to sustain macroscopic quantum tunneling\cite{longobardi2011A} at lower temperatures.

\section{Theoretical Background}

The significance of the analysis of phase diffusion phenomena extends to the more general problem of the motion of a Brownian particle in a periodic potential, which is behind the Resistively Shunted Junction (RSJ) model for the dynamics of the phase difference across a Josephson junction \cite{barone,likharevbook}. It can describe many different physical phenomena, i.e. transport on crystalline surfaces\cite{Pollak_1993}, rotating dipoles in external fields\cite{reguera_2000}, charge density waves\cite{gruner_1981}, and particle separation by electrophoresis\cite{Ajdari1991}. Kramers\cite{kramers} first studied the one-dimensional escape dynamic of a quantum Brownian particle in presence of damping, in the low and high damping limits, and subjected to a tilted potential $U(z)$
\begin{equation}
U(z) = V(z) - Fz
\end{equation}
where $V(z) = V(z + P)$ denotes a periodic potential of period P and F is
an external static force.
\begin{figure}
\includegraphics[width=6.5cm]{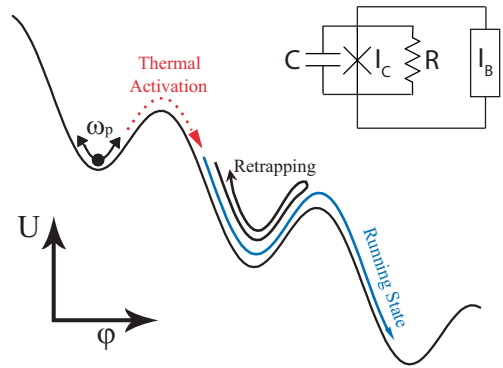}
\caption{Tilted periodic potential and the equivalent circuit of the RSJ model.}
\label{fig:potential}
\end{figure}
The dynamics of the phase difference $\varphi$ of the superconducting order parameter across the junction of a current-biased JJ is equivalent to the Brownian motion of a particle of "mass" C (the junction capacitance) in a tilted periodic potential\cite{barone,likharevbook}
\begin{equation}
U(\varphi)=-E_J\left(\cos\varphi +\frac{I}{I_c}\varphi\right)
\end{equation}

In this case the external static force is controlled by the bias current I and is given by $E_J (I/I_c)$ where $E_J=\hbar I_c / 2 e$ is the Josephson energy. The motion of the particle along the potential is also subject to friction, whose strength can be characterized by the above mentioned junction quality factor Q. In this model the superconducting branch of the junction current-voltage (I-V) characteristic corresponds to the confinement of the particle in one well of the potential. The escape from this metastable state corresponds to the appearance of a finite voltage across the junction. As it is shown in fig. \ref{fig:potential}, in case of low damping the escaped particle gains sufficient energy to roll down the potential in the so-called running state, meanwhile if the damping is sufficiently high, escape due to thermal hopping does not necessarily lead to runway down the tilted potential\cite{kivioja2005,mannik2005}. Following an event of escape the particle may travel down the potential for a few wells and then be retrapped in one of the following minima of the potential\cite{martinis_kautz}. At low bias the process of escape and retrapping may occur multiple times generating extensive diffusion of the phase until an increase of the tilt of the potential, due to a change in the bias current, raises the velocity of the particle and the junction can switch to the running state\cite{fenton2008}.

This phenomenon of phase diffusion is more evident when studying the temperature dependance of the switching probability\cite{fulton1974}:
\begin{equation}
P(I)=\frac{\Gamma (I)}{dI/dt} \exp {\left[-\int_{0}^{I} \frac{\Gamma (I')}{dI'/dt} dI'\right]}
\label{eq:4.4}
\end{equation}
where the rate of escape due to thermal activation is given by Kramers formula\cite{kramers}
\begin{equation}
\Gamma_{TA}(I)= a_t \frac{\omega_p(I)}{2\pi}\exp \left(-\frac{\Delta U(I) }{k_B T}\right)
\label{eq:TA}
\end{equation}
being $\Delta U$ the height of the energy barrier between consecutive potential wells and the prefactor is $a_t=4/[(1+Q k_b T / 1.8 \Delta U)^{1/2}+1]^2$.
For underdamped junctions the width $\sigma$ of the switching distributions has a monotone dependence on temperature as $\sigma \propto T^{2/3}$. In the case of moderately damped junctions, due to the retrapping process the switching dynamic is modified; below a turn-over temperature $T^*$ the width of the switching distributions follows the usual $\sigma \propto T^{2/3}$ , while for $T>T^*$, $\sigma$ is reduced with increasing temperature. In this paper we also report on the observation of this behavior which we fit using the FW model.

\section{Samples Fabrication Process and Experimental Setup}
In NbN/MgO/NbN JJs a 1 nm thick barrier provides $J_c$ of about $3 A/cm^2$, which is the lowest value ever reported for  NbN based junctions. For circular junctions with a diameter of $10 \mu m$, $I_c$ is about 1-2 $\mu A$ and falls under the criteria of the moderately damped regime, as extensively discussed below. In the trilayer, epitaxially grown at ambient temperature on a single-crystal MgO substrate \cite{kawakami_jap}, the NbN base (BE) and counter electrode (CE) are both 200nm thick and were deposited using DC magnetron sputtering with a Nb target in a mixture of 5 parts argon and 1 part nitrogen gas. The MgO barrier is about 1.0nm thick and was deposited by rf sputtering. This step was followed by a reactive ion etch (RIE) for junction definition and by the deposition of a MgO insulating layer patterned by a lift-off process. The process is concluded with the deposition of a 350nm NbN wiring layer which was patterned and defined by RIE. The realized junctions have a superconducting transition temperature of about 16.6 K for both electrodes. A more detailed description of the fabrication process can be found in a paper from Kawakami et al.\cite{kawakami_jap}.

\begin{figure}
\includegraphics[width=6.5cm]{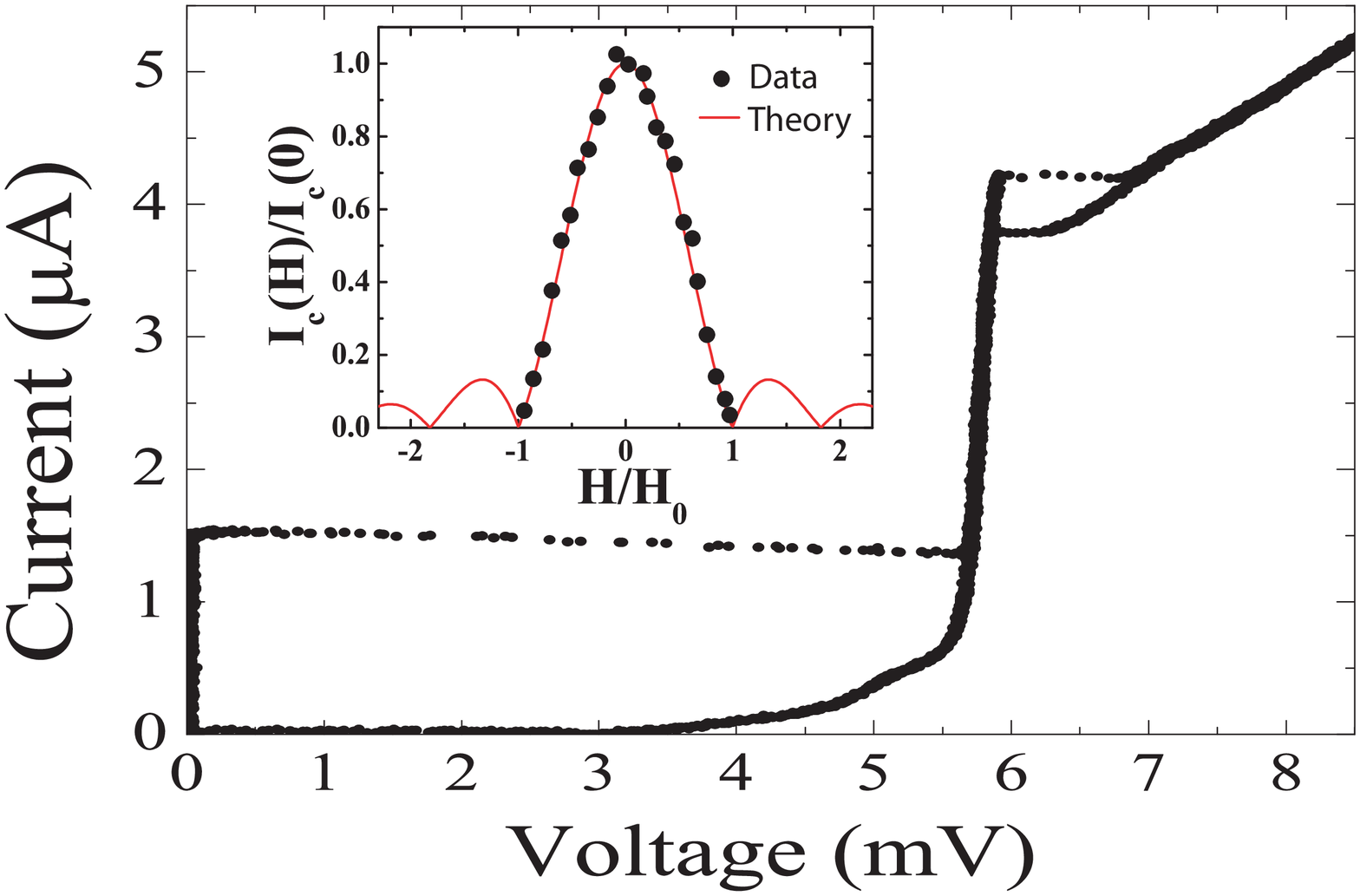}
\caption{Current-Voltage characteristic of a $10 \mu m$ diameter JJ measured at 290mK. The inset shows the magnetic field dependence of the Josephson tunnel current, with magnetic field values normalized to the first minimum.}
\label{fig:fab}
\end{figure}

The junctions have been tested through measurements of the I-V characteristics (see fig. \ref{fig:fab})and of switching current distribution (see next section). From the magnetic field dependence of the critical current (shown in the inset of fig. \ref{fig:fab}) we estimated the London penetration depth at 300mK to be about $\lambda_L=190nm$ which is in good agreement with previously measured values for epitaxially grown NbN\cite{tu,komiyama1996}, therefore for a barrier of thickness t=1 nm the Josephson  penetration depth turns out to be\cite{barone} $\lambda_J = (\hbar c^2 / 8\pi e J_c d)^{1/2} = 150 \mu m$, where $d=2\lambda_L + t$.

\begin{figure}
\includegraphics[width=6.5cm]{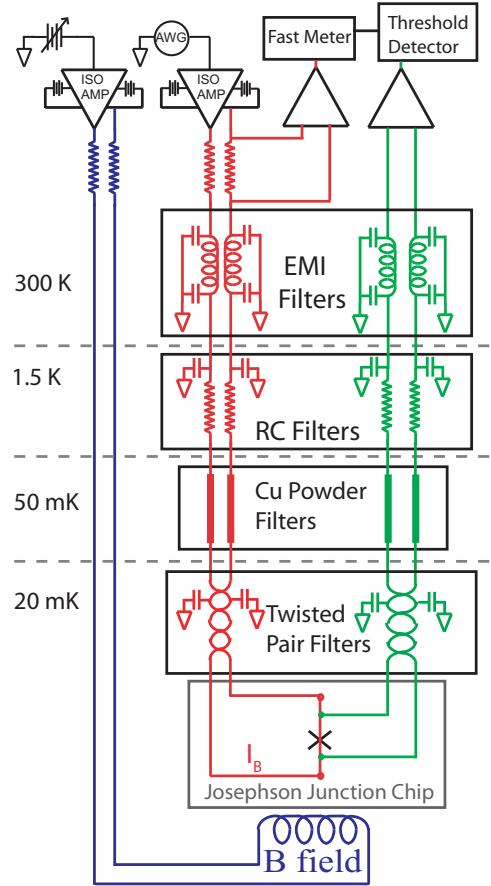}
\caption{(color online) Schematic of measurement electronics including thermal anchoring and the
various stages of filtering.}
\label{fig:circuit}
\end{figure}

To study the escape rates of a NbN/MgO/NbN Josephson junction we have thermally anchored the sample to the mixing chamber of a He3/He4 Oxford dilution refrigerator and performed measurements of the junction switching current probability.
Figure \ref{fig:circuit} shows a block diagram of the experimental setup including  room temperature electronic and filtering.
The room temperature circuits have been optimized to minimize the effect of unwanted noise. In order to avoid ground loops and noise pick up, the whole
experiment is designed to have a single ground \cite{ott,morrison}.
All connections to the chip are floating and are only capacitively
coupled to ground through the filters. All grounded signal sources pass
through battery powered unity gain isolation amplifiers that
effectively disconnects this signal from the earth ground.  The current paths are all designed to be symmetric with respect to the chip to reduce the effect of
common mode noise \cite{bennett2009}.
This allows for sufficient decoupling from the ground while keeping
the amplifier from saturating due to charging by providing a return
current path. The amplifiers are designed to have 100 $M \Omega$ of resistance between their inputs and the common of the battery circuit. All signals entering the fridge are isolated,
shielded and filtered allowing the dilution refrigerator itself to
act as an rf-shield for the cold portion of the experiment. In our system we use a room temperature electromagnetic interference filter stage
followed by low pass RC filters with a cut-off frequency of 1.6MHz anchored at 1.5 K. Further filtering is provided by a combination of copper powder \cite{milliken:024701} and twisted pair filters \cite{spietz} thermally anchored at the mixing chamber of the dilution refrigerator.

\section{Data and Analysis \label{data}}
\begin{figure}
\includegraphics[width=6.5cm]{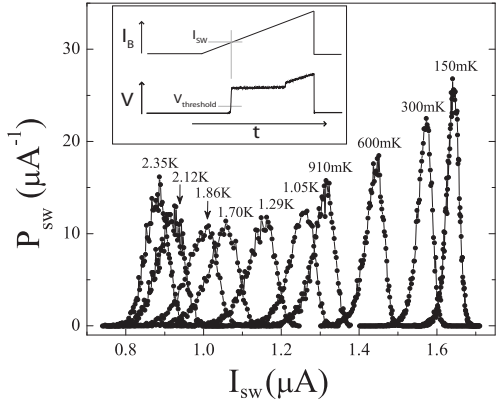}
\caption{Switching current probability distribution at $B=0G$ for different bath temperatures. The inset shows the signal sequence used to acquire the data.}
\label{fig:distribution}
\end{figure}

\begin{figure}
\includegraphics[width=6.5cm]{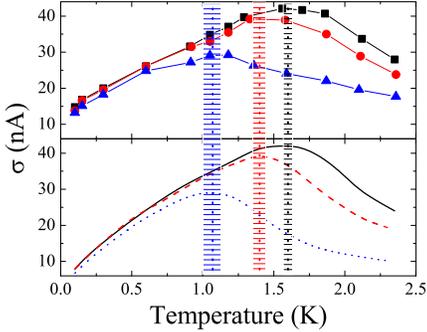}
\caption{Top frame: temperature dependence of the standard deviation, $\sigma$, of the switching distributions for $B=0G$ (squares), $B=3.05G$ (circles), and $B=6.09G$ (triangles). Bottom frame: a numerical simulation of the data. Vertical dotted lines have been inserted in correspondence of the values of $T^*$ including error bars. Data and numerical simulations are in good agreement in the whole temperature range and for all magnetic fields within error bars..}
\label{fig:sigma}
\end{figure}

The signal sequence used to measure the switching current distribution (SCD) is shown in the inset of fig.\ref{fig:distribution}. The bias current of the junction is ramped at a constant sweep rate $dI/dt=122 \mu A/s$, the voltage is measured using a low noise differential amplifier and is fed into a threshold detector which is set to generate a pulse signal when the junction switches from the superconducting state to the finite  voltage state. This signal is used to trigger a fast volt meter to record the value of the switching current \cite{doug2007}. This procedure is repeated at least $10^4$ times at each temperature, which allows to compile a histogram of the switching currents. In fig.\ref{fig:distribution}, we report the SCD curves collected over a wide range of temperatures in absence of an externally applied magnetic field. Distinctive fingerprints of phase diffusion, due to multiple hopping and retrapping, can be found in the temperature dependence of the width $\sigma$ of the SCD curves, which  is shown in Fig. \ref{fig:sigma}. The most striking effect observable in Fig.\ref{fig:sigma} is the appearance of an anticorrelation between the temperature and the width of the switching distributions\cite{kivioja2005,mannik2005,krasnov2005}. At low temperatures the $\sigma$ follow the expected $T^{2/3}$ dependence, deviations are evident in proximity and above a  "critical temperature" $T^*$ where the temperature derivative of $\sigma(T)$ becomes negative. Experimental data (upper frame in Fig.\ref{fig:sigma}) are well reproduced by the expected values (lower frame in Fig.\ref{fig:sigma}), calculated  on the basis of the physical arguments of phase diffusion.

The simulations shown in Fig.\ref{fig:sigma} in particular, are based on the recent work on phase diffusion by Fenton and Warburton \cite{fenton2008}. The phase difference $\varphi(t)$ is a solution of the following Langevin differential equation:
\begin {equation}
\varphi_{tt}+\varphi_t/Q+\iota+\iota_N=0
\label {lang}
\end {equation}
Times t are normalized to $1/\omega_p$; $\iota$ is the bias current normalized to critical current $I_{co}$ and $\iota_N$ is a Gaussian correlated thermal noise current, i.e.:
$$
\left<\iota_N(t),\iota_N(t')\right>=\sqrt{2\pi k_B T/QI_{co}\Phi_o }\delta(t-t').
$$
Stochastic dynamics is simulated by integrating the above Langevin equation by a Bulirsh-Stoer integrator using as noise generator the cernlib routine RANLUX \cite{notasim}. Simulations have been carried out for different temperatures and dissipation values.

The multiplicity of switching modes between the running and the trapped states raises a problem of how to define an escape event. In our simulations the condition to define the switch is $V(\iota,T)\ge V(\iota,0)/2$, where $V$ represents the average velocity of the phase-particle in the washboard potential. In other words the particle spends in the running state more than  50\% of observation time. Typical runs for simulations of Eq.(\ref{lang}) will last from $4\cdot10^6$ to $6\cdot10^6$ normalized time units, i.e., $6\cdot10^5$ to $9\cdot10^5$ plasma periods. Observation time for each point generated in the I-V characteristics is $2\cdot10^4$ time units, which is a long enough time to ensure that the average time spent in running/zero voltage state does not vary as a function of the observation time\cite{kautz90}.
To obtain the SCD we have simulated a number of escape events between $3000$ and $5000$, which is similar to the number of counts experimentally measured.
\begin{figure}
\includegraphics[width=6.5cm]{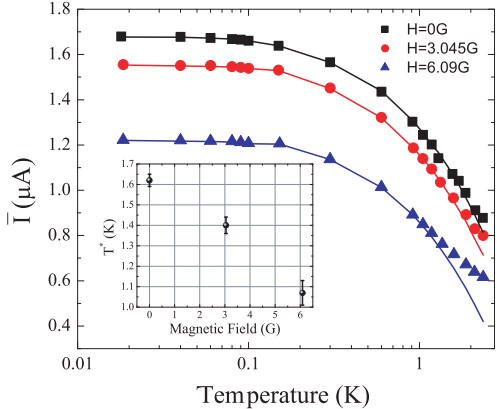}
\caption{Temperature dependence of mean current, $\overline{I}$, of the switching distributions for $B=0G$ (squares), $B=3.05G$ (circles), and $B=6.09G$ (triangles). The solid lines are the calculated values of $\overline{I}$ considering thermal activation and no retrapping. The inset shows the value of $T^*$ vs the applied magnetic field.}
\label{fig:I}
\end{figure}
Simulated curves of $\sigma$ vs T for different values of the magnetic field are plotted in Fig. 5. The magnetic field works as a knob to tune $T^*$ and provides an additional validity test for the estimate of $Q= 2.7 \pm 0.1$. From fitting of the SCD and of the moments of the distributions at temperatures below $T^*$ we have estimated the value of the zero temperature critical current to be $I_{co}=1.91 \pm 0.03  \mu A$. We have also estimated values for the junction capacitance and plasma frequency and obtained $C=0.3 pF$ and $\omega_p \simeq$ 22 GHz \cite{longobardi2011A}.

The quality of the fitting procedure is even more significant if we consider that we do not have any degree of freedom associated to a possible frequency dependence of Q as occurring in other experiments \cite{kautz90,mannik2005,Bae2009}. For instance the procedure used for our analysis is different from that used by M\"{a}nnik et al. \cite{mannik2005}, where retrapping probability is calculated independently for different dissipations. An analysis of full escape rate is made by combining the $0\rightarrow V$ escape probability with retrapping probability which allows to extract the resistance by fitting the escape rate curves.

In Fig. \ref{fig:I} the mean values $\overline{I}$ of the SCD (data points) are plotted along with the expected values (solid lines) without taking into account retrapping effects. Due to the onset of retrapping events it is necessary to provide a larger tilt to the energy potential to allow the system to switch to the running state. Discrepancies at higher T demonstrate that  the experimental values of $\overline{I}$ at higher temperatures (above $T^*$) are greater than the predicted values, which only consider the effects of thermal activation. Due to the dependence of $T^*$ with the external magnetic field, as shown in the inset of Fig. \ref{fig:I}, this latter effect is particularly evident for the data taken in presence of a 6.09G static magnetic field, where the onset of retrapping occurs at a lower temperature.

\begin{figure}
\includegraphics[width=6.5cm]{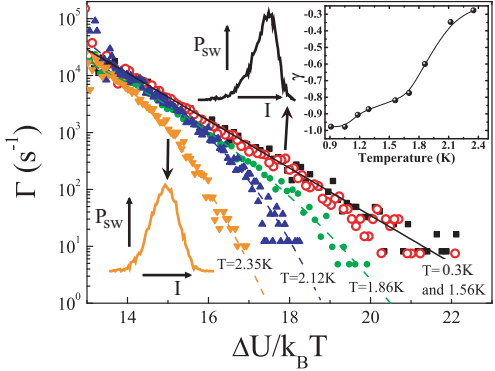}
\caption{Escape rates (symbols) as a function of the barrier height at zero magnetic field, for temperatures near $T^*$, and numerical simulations (dashed lines). Below $T^*$ the fit has been calculated using Kramers formula for thermal activation (solid line). The inset shows the experimental value of the skewness of the switching distribution.}
\label{fig:rates}
\end{figure}

Phase diffusion also appears in the escape rates $\Gamma$, shown in figure \ref{fig:rates}  as a function of the ratio between the barrier height and the thermal energy. The escape rates are calculated from the switching distributions using eq. \ref{eq:4.4}. In the thermal activation regime the distributions are asymmetric and skewed to the left, and $\Gamma$ values all fall onto the same line, as it is the case for the reported data from T=0.3K to 1.56K. Retrapping processes cause a progressive symmetrization of the switching distribution, as it can be seen from the inset in the bottom left corner of Fig. \ref{fig:rates}, and a bending in the $\Gamma$ vs $u=\Delta U/k_B T$.
We use the same procedure previously described to evaluate the numerically simulated escape rates $\Gamma$ as function of reduced barrier height $u$. Numerical data have been obtained by a polynomial fit of numerical escape rates in order to compare it with experiments. The same value of the Q factor is obtained by fitting the $\Gamma(u)$ curves, shown in dashed lines in Fig. \ref{fig:rates}.

The symmetrization of the switching distribution due to the interplay between escape and retrapping events can be clearly observed by plotting, as a function of temperature, the skewness of the distributions $\gamma$\cite{Note1}. For the data with no external magnetic field, we report such plot in the inset of fig. \ref{fig:rates}. For the lowest temperatures we obtain $\gamma=-1$, which is consistent with the case of switching current distributions in the quantum or thermal regime. As the temperature increases the distributions become more and more symmetric as $\gamma$ tends to zero. It should be noted that for these data the temperature $T^*$ at which the derivative of $\sigma(T)$ changes sign is equal to $1.62\pm0.03$ K and that the skewness starts increasing already at about $1.2K$, which is a clear indication that the onset of retrapping phenomena occurs well below $T^*$.\cite{fenton2008}

\section{Discussion and Concluding Remarks}

\begin{table*}
\caption{\label{table}Comparison of device parameters.}
\begin{ruledtabular}
\begin{tabular}{ccccccccc}
\textbf{Author}  & Device Structure  &   R ($\Omega$)     &     $I_{co}$ ($\mu A$)  & Area ($\mu m^2$)    &       $Q(I=0)$     &        $\Delta U(\overline{I})/k_B T^*$ & $E_J/E_C$\\ \hline \\
This work & NbN/MgO/NbN JJ  &  65 & 1.91 & 78.5 & 2.7 & 17 & 14760 \\
\emph{Kivioja et al.} \cite{kivioja2005} & Al/AlOx/Al dc SQUID  &  500 & 0.2 & 1.0 & 3.9 & 14 & 515 \\
\emph{Kivioja et al.} \cite{kivioja2005}  & Al/AlOx/Al JJ &  230 & 0.63 & 2.6 & 3.6 & 18 & 2110 \\
\emph{M\"{a}nnik et al.} \cite{mannik2005} & Nb/AlOx/Nb dc SQUID  &  70 & 4.25 & 1.0 & 2.4 & 15 & 9850 \\
\emph{M\"{a}nnik et al.} \cite{mannik2005} & Nb/AlOx/Nb dc SQUID  &  70 & 2.9 & 2.9 & 3.3 & 17 & 19420 \\
\emph{Bae et al.} \cite{Bae2009} \footnote{In this paper the authors estimated the fit parameters to be temperature dependent. Here we report the values at the lowest experimental temperature T=1.5K.}  & Bi-2212 Intrinsic JJ &  62 & 1.26 & 7.3 & 2.2 & 14 & 10710 \\
\emph{Yu et al.} \cite{Yu_2011} & Nb/AlOx/Nb JJ  &  1800  & 0.122 & 0.4 & 4.8 & NA & 62\\
\emph{Yu et al.} \cite{Yu_2011}  & Nb/AlOx/Nb JJ  &  315 & 0.48 & 1.5 & 3.3 & 12 & 950\\
\end{tabular}
\end{ruledtabular}
\end{table*}

A change in the sign of the derivative of the second moment of the distribution at a turn-over temperature $T^*$ and a modification of the shape of the distributions at temperature around $T^*$ are robust signatures of the phase diffusion regime, and also occur in our NbN/MgO/NbN junctions, as discussed in the previous section. A non exhaustive list of Josephson devices (including ours) that have displayed a similar PDR behavior is reported in table \ref{table} along with the most relevant device parameters.

The papers on which table \ref{table} is based report on similar experimental results but their interpretation differs in few assumptions, as properly pointed by Fenton and Warburton \cite{fenton2008} . For instance Kivioja et al.\cite{kivioja2005} interpreted their results within the semiclassical model of Larkin and Ovchinnikov \cite{larkin:185,larkin:1060}Since in dc-squids there are few energy levels and the hypothesis of continuous energy spectrum is not valid, they used a model which takes account both phase diffusion and level quantization. On the other hand this model, which assumes separated levels in the metastable well, is not properly valid for a single Josephson junction since the number of energy levels is large and the separation is smaller than their width. M\"{a}nnik et al.\cite{mannik2005} and Bae et al.\cite{Bae2009} calculated the retrapping probability through Monte Carlo simulations and included frequency dependent damping. The authors expressed the net escape rate as a sum of probabilities of multiple escape-retrapping events based on thermal escape rate and retrapping probability. The probability of retrapping is considered as a time-independent quantity which is in contrast with the work of Ben-Jacob et al.\cite{ben-jacob1982} in which retrapping is modeled by a rate and therefore by a probability increasing proportionally to the time spent in the running state. The very good fitting of experimental curves obtained in this work using the FW approach\cite{fenton2008} confirms the occurrence of a multiple-retrapping regime with a large number of escapes of duration of $\Gamma_R^{-1}$, and in particular it confirms that the time dependence of the retrapping probability cannot be ignored. If the fast scattering time $\tau$ plays a role in diffusive process is a topic of further investigations.

As it can be seen from table \ref{table}, our experiment confirms that, independently of the physical size of the device, all the junctions exhibiting phase diffusion over a large range of materials and geometry have a low critical current, $2<Q<5$ and $12<\Delta U(\overline{I})/k_B T^*<18$, which are therefore the relevant parameters signaling the insurgence of multiple escape and retrapping in a washboard potential.

The possibility to have extremely low critical current density can be functional to investigate phase dynamics at extreme conditions. An example is given by a recent experiment on submicron Nb/AlOx/Nb junctions \cite{Yu_2011}. Data show an anomalous $\sigma (T)$ dependence with a negative $d\sigma/dT$ over the entire temperature range and a saturation at low temperatures. This regime can be achieved by engineering junctions with low critical current, such that the turn over temperature $T^*$ is lower or comparable to the quantum crossover temperature $T_{cross}$\cite{devoret1985}. In this case the enhancement of $\sigma$ when increasing the temperature, characteristic of the thermal regime, is not observed. Junctions with intrinsically low critical current density, such as the one reported in the present work, could represent an interesting term of comparison to study these kind of unconventional regimes using standard micrometer junctions.

In conclusions we have proved that low $J_c$ NbN/MgO/NbN JJs are characterized by a transition from thermal activation regime to phase diffusion. This is consistent with what has been observed in other types of junctions with similar values of Q and $\Delta U(\overline{I})/k_B T^*$. The experimental results are well described by a numerical model involving a frequency independent damping which demonstrates an efficient way to estimate the dissipation in moderately damped JJs.

\section{acknowledgments}
This work was supported in part by STREP "MIDAS-Macroscopic Interference Devices for atomic and Solid State Physics: Quantum Control of Supercurrents" and by a Marie Curie International Reintegration Grant number 248933 "hybMQC" within the 7th European Community Framework Programme. We also acknowledge partial support by MIUR PRIN 2009 under the project "SuFET based on nanowires and HTS" and by ISCRA the Italian SuperComputing Resource Allocation through grant IscrB\_NDJJBS 2011.

\end{document}